# Quadrefringence of optical vortices in a uniaxial crystal

T. Fadeyeva[1], A. Rubass[1], Yu. Egorov[1], A. Volyar[1*], G. Swartzlander Jr[2]

[1]Physical Department, Taurida National University, Simferopol 95007, Crimea Ukraine
[2]Optical Science Center, University of Arizona, Tucson, Arizona, 85721
*Corresponding author: volyar@crimea.edu

**Abstract:** The splitting of a single optical vortex into four separate ones in a singular beam is theoretically and experimentally described for the propagation of light obliquely through a uniaxial crystal. Also we found the condition under which the new-born vortices in each four individual beams propagate independently without dislocation reactions and have different locations in all beams for any crystal lengths.

## I. Introduction

The predictions of a new branch of modern optics called singular optics [1] sometimes enable us to catch sight of utterly unexpected features of well-known physical phenomena. For example, Berry recently demonstrated the conical refraction of singular beams in biaxial crystals [2]. The vortex nature of light beams propagating along the crystal optical axis was found to be connected with the conical singularity of the dispersion surface and the structure of the Poggendorff rings.

In this paper we illustrate an unexpected phenomenon for beams in a uniaxial birefringent medium. Here we demonstrate that an optical vortex nested in a simple singular beam propagating obliquely to the crystal optical axis may emerge as four independent optical vortices imprinted in four individual beams of the complex, combine beam field. This phenomenon may be called *vortex-quadrefringence* (from quadru-refringence). The necessary conditions for the generation of such a vortex splitting are theoretically determined and experimentally verified. Furthermore we propose the resolution criterion whereby the four vortex-beams become distinguishable in the experiment.

The vortex propagation in uniaxial and biaxial crystals has been discussed early in a lot of papers [3-11] where a main attention was focused as a rule on propagating a fundamental Gaussian beam along a crystal optical axis. In a series of recent papers [12-14] some main features of oblique singular beams has been outlined. Conditions for generation of singly-charged vortex-beams were treated in Ref. [12], whereas authors of Ref. [13, 14] found and analyzed a complicated network of polarization singularities occurring in combine oblique beam. However they have considered the processes for the field parameters defined far from the condition of the beam splitting and, consequently, far from the subject of our discussion.

The aim of this paper is to trace theoretically and experimentally the development of the process of splitting the initial optical vortex into four ones imprinted in individual crystal-propagating beams and bring to light the conditions under which the combine singular beam can be emerged as four separate vortex-beams.

The paper is organized as follows. In Section 2 we obtain the set of the off-axis paraxial beams with "mismatched" optical vortices and deduce the field expressions for an incident off-axis fundamental vortex beam. In section 3 we consider intrinsic features of dislocation reactions in the field components via variations of the beam inclination to the crystal axis. Section 4 is devoted to analysis of the indistinguishability limit that restricts the independent propagation of the individual beams that have splintered. Section 5 shows that the nonuniform distribution of polarization states in the vicinity of polarization singularities is an inherent property – even when strongly splintered beams weakly overlapping.

## II. The solutions to the paraxial wave equation

We consider the propagation of a light beam whose monochromatic electric field may be expressed as $\mathbf{E}(\mathbf{r},t) = \mathbf{E}(\mathbf{r})\exp(-i\omega t)$, where $\mathbf{r} = \hat{\mathbf{x}}x + \hat{\mathbf{y}}y + \hat{\mathbf{z}}z$. We will be interested in the fundamental vortex-beam whose complex amplitude at the plane $z' = 0$ inside the crystal has a form:

$$\mathbf{E}(x', y', z' = 0, \alpha_o) = \mathbf{A}(\alpha_o)[(x' - iy')/w_0]\exp(-r_\perp'^2/w_0^2) \quad (1)$$

in the reference frame $\{x', y', z'\}$ tilted in the $(x, z)$ plane to the crystal optical axis at the angle $\alpha_o$. Here, $r'^2_\perp = x'^2 + y'^2$, $w_0$ is a beam waist radius. The beam is assumed to propagate through a uniaxial crystal with a permittivity tensor written in a reference frame $\{x, y, z\}$ of the crystal as

$$\hat{\varepsilon} = \begin{pmatrix} \varepsilon_1 & 0 & 0 \\ 0 & \varepsilon_2 & 0 \\ 0 & 0 & \varepsilon_3 \end{pmatrix}, \quad (2)$$

so that $\varepsilon_1 = \varepsilon_2 = \varepsilon$, $\varepsilon$ and $\varepsilon_3$ are real constants, and $\varepsilon_3 < \varepsilon$. Here the crystal $c$-axis is parallel to the $z$-axis of our system. As well known [15], an oblique beam in a uniaxial crystal gets elliptically deformed. However, we will consider a very small inclination angles $\alpha$ in the crystal such that $\alpha \ll 1$: $\sin\alpha \approx \alpha$, $\cos\alpha \approx 1 - \alpha^2/2$ and this geometrical deformation does not manifest itself in an explicit form. In the crystal the complex amplitude $\mathbf{E}(x, y, z > 0)$ obeys the wave equation

$$\left(\nabla^2 + k^2 \hat{\varepsilon}\right) \mathbf{E} = \nabla(\nabla \cdot \mathbf{E}) \quad (3)$$

where $k = 2\pi/\lambda$ is the vacuum wavenumber and $\lambda$ is the vacuum wavelength.

We will restrict our attention to a paraxial approximation. We represent the beam as if its wavefront propagates along the $z$-axis with a wave vector of magnitude $k_o = k\sqrt{\varepsilon}$ while the transverse amplitude of the field is:

$$\mathbf{E}_\perp(x, y, z, \alpha) = \tilde{\mathbf{E}}_\perp(x, y, z, \alpha) \exp(-i k_o z) \quad (4)$$

where $\tilde{\mathbf{E}}_\perp(x, y, z, \alpha_o)$ is a slowly varying amplitude having components in the transverse $x,y$-plane. The inclination of the beam axis we will take into account with a shift of the origin along the $y$-axis in imaginary region [16] at the distance $y_0 = -\alpha z_0$: $y \to y + i\alpha z'_0$, where $z'_0 = k' w_0^2 / 2$, $k'$ is a wavenumber of the beam in the crystal. In paraxial approximation, we come to the wave equation [11, 17]:

$$\left(\nabla_\perp^2 + 2 i k_o \partial_z\right) \tilde{\mathbf{E}}_\perp = \frac{\Delta\varepsilon}{\varepsilon_3} \nabla_\perp \left(\nabla_\perp \cdot \tilde{\mathbf{E}}_\perp\right) \quad (5)$$

where $\nabla_\perp \cdot \mathbf{E}_\perp = \partial_x E_x + \partial_y E_y$. Since we will not consider $E_z$ component in the paper, hereafter we stop using the subscript ($\perp$) for the field.

The vector equation (5) can be reduced to a simple scalar form by means of two substitutions:

$$\tilde{\mathbf{E}}^{(o)} = \left(\hat{\mathbf{x}} \partial_y - \hat{\mathbf{y}} \partial_x\right) \Psi_o \quad (6)$$

and

$$\tilde{\mathbf{E}}^{(e)} = \nabla_\perp \Psi_e . \quad (7)$$

where $\Psi_o(x, y, z, \alpha_o)$ and $\Psi_e(x, y, z, \alpha_e)$ are arbitrary scalar fields that must satisfy the paraxial conditions for $\tilde{\mathbf{E}}^{(o)}$ and $\tilde{\mathbf{E}}^{(e)}$. Eqs. (6) and (7) describe groups of o*rdinary* and *extraordinary* beams with two values of wavenumbers $k' \to k_o, k_e$ $z'_0 \to z_o, z_e$ and two values of angles $\alpha \to \alpha_o, \alpha_e$ whereas we suppose that waist radii of the ordinary $w_o$ and the extraordinary $w_e$ beams are the same $w_o = w_e = w_0$. Notice that eqs (6) and (7) in common with the condition (1) represent single-valued functions. Thus, eq. (5) transforms into two scalar equations:

$$\left(\nabla_\perp^2 + 2 i k_o \partial_z\right) \Psi_o = 0, \quad (8)$$

$$\left(\nabla_\perp^2 + 2 i k_e \partial_z\right) \Psi_e = 0 \quad (9)$$

with $k_e = (\varepsilon_3/\sqrt{\varepsilon}) k_0$. Particular the simplest solutions of these equations are

$$\tilde{\Psi}_o = \sigma_o^{-1} \exp\left\{-(x^2 + y_o^2)/w_0^2 \sigma_o\right\} \times \\ \times \exp\left(-\alpha_o^2 k_o z_o / 2\right) \quad (10)$$

$$\tilde{\Psi}_e = \sigma_e^{-1} \exp\left\{-(x^2 + y_e^2)/w_0^2 \sigma_e\right\} \times \\ \times \exp\left(-\alpha_e^2 k_e z_e / 2\right) \quad (11)$$

where $y_o = y + i\alpha_o z_o$, $y_e = y + i\alpha_e z_e$, $z_o = k_o w_0^2/2$, $z_e = k_e w_0^2/2$, $\sigma_o(z, k_o) = 1 - iz/z_o$, and $\sigma_e(z, k_e) = 1 - iz/z_e$. Here the beam waists, both of radial size $w_0$, coincide with the plane $z = 0$. Besides, with the help of eq. (1) we find $\tilde{\Psi}_o(z = 0) = \tilde{\Psi}_e(z = 0)$ so that $\alpha_o k_o = \alpha_e k_e$.

Combining eqs. (4), (6), (7), (10) and (11) we obtain particular solutions for the ordinary and extraordinary fundamental vortex beam fields in the circularly polarized basis: $\hat{\mathbf{e}}_+ = \hat{\mathbf{x}} + i\hat{\mathbf{y}}$, $\hat{\mathbf{e}}_- = \hat{\mathbf{x}} - i\hat{\mathbf{y}}$ in the form

$$\mathbf{E}^{(o)} = \tilde{\mathbf{E}}^{(o)} \exp(-i k_o z) = \\ = \left\{\hat{\mathbf{e}}_+ \frac{x - i y_o}{\sigma_o(z, k_o)} - \hat{\mathbf{e}}_- \frac{x + i y_o}{\sigma_o(z, k_o)}\right\} \frac{\tilde{\Psi}_o}{w_0} \exp(-i k_o z), \quad (12)$$

$$\mathbf{E}^{(e)} = \tilde{\mathbf{E}}^{(e)} \exp(-i k_o z) = \\ = \left\{\hat{\mathbf{e}}_+ \frac{x - i y_o}{\sigma_e(z, k_e)} + \hat{\mathbf{e}}_- \frac{x + i y_o}{\sigma_e(z, k_e)}\right\} \frac{\tilde{\Psi}_e}{w_0} \exp(-i k_o z), \quad (13)$$

Clearly each polarization component of the beams in eqs. (12) and (13) contains a fundamental vortex. The right-handed polarized components have negative topological charges, $l_+^{(o)} = l_+^{(e)} = -1$, that are displaced along x-axis to the point $x_+^{(o)} = x_+^{(e)} = -\Delta x_o$. The left-handed polarized components have positive charges, $l_-^{(o)} = l_-^{(e)} = +1$, and displacements $x_-^{(o)} = x_-^{(e)} = \Delta x_o$, where $\Delta x_O = |\alpha_o z_O|$. An intriguing property of eqs. (12) and (13) is that the vortex trajectories are parallel to the z-axis whereas the beam envelopes deviate from the z-axis by the angles $\alpha_o$ and $\alpha_e$, respectively.

Clearly eqs. (15, 16) can not satisfy eq.(1) for any polarization states $\mathbf{A}(\alpha)$. Let us at first to soften this requirement demanding one of the circularly polarized components of the sum of eqs (15, 16) to be zero at the plane z=0, say, a left-handed polarized component vanishes: $E_-(x, y, z = 0)$. As a result we find:

$$\mathbf{E}_1 = \tilde{\mathbf{E}}^{(o)} + \tilde{\mathbf{E}}^{(e)} = \hat{\mathbf{e}}_+ \frac{x - i y_o}{w_0} \left( \frac{\tilde{\Psi}_o}{\sigma_o} + \frac{\tilde{\Psi}_e}{\sigma_e} \right) - \hat{\mathbf{e}}_- \frac{x + i y_o}{\sigma_o} \left( \frac{\tilde{\Psi}_o}{\sigma_o} - \frac{\tilde{\Psi}_e}{\sigma_e} \right)$$ (14)

In fact, it means that if we form the field: $\mathbf{E}_1(z=0) = \hat{\mathbf{e}}_+ (x - i y_o)\tilde{\Psi}^{(o)}(z=0)/w_0$ at the plane z=0 of the crystal whose vortex is shifted relative to the origin, the energy flux will evolve at some angle to the z-axis while the optical vortex will propagates parallel to z-axis.

In order to satisfy eq.(1) let us find one more field lacking in phase singularities similar to that of a fundamental Gaussian beam in free space. Towards this end we make use of the following recipe [11]: $\tilde{\mathbf{G}} = \int_0^z \tilde{\mathbf{E}}_1(x, y, z)dz$. After a little algebra we come to the expression:

$$\tilde{\mathbf{G}} = \hat{\mathbf{e}}_+ (\tilde{\Psi}_o + \tilde{\Psi}_e) - \hat{\mathbf{e}}_- \left( \frac{x + i y_o}{r_o} \right)^2 \times$$
$$\times \left[ \frac{w_0^2}{r_o^2} (\sigma_o \tilde{\Psi}_o - \sigma_e \tilde{\Psi}_e) + (\tilde{\Psi}_o - \tilde{\Psi}_e) \right]$$ (15)

where $r_o^2 = x^2 + y_o^2$. The above expression characterizes evolution of a Gaussian beam in a crystal [3, 11]. Notice that a left-handed component $G_-$ of the oblique beam carries over double-charged optical vortex that does not spread together with energy flux.

Superposition of a singular beam with a shifted vortex (14) and a Gaussian beam (15): $\tilde{\mathbf{E}} = \tilde{\mathbf{E}}_1 - \bar{\alpha}\tilde{\mathbf{G}}$ (where $\bar{\alpha} = \alpha_o z_o / w_0$) results in centering the optical vortex at the plane z=0 and fulfilling the condition (1) for a right-handed polarized component $E_+$ whereas a left-handed polarized component vanishes $E_-(z=0) = 0$:

$$\tilde{E}_+ = \left\{ \frac{x - i(y - \alpha_o z)}{w_0 \sigma_o} \tilde{\Psi}_o + \frac{x - i(y - \alpha_e z)}{w_0 \sigma_e} \tilde{\Psi}_e \right\},$$ (16)

$$\tilde{E}_- = \frac{x + i y_o}{w_0} \left[ \frac{\tilde{\Psi}_o}{\sigma_o} - \frac{\tilde{\Psi}_e}{\sigma_e} \right] + \bar{\alpha} \left( \frac{x + i y_o}{r_o} \right)^2 \times$$
$$\times \left[ \frac{w_0^2}{r_o^2} (\sigma_o \tilde{\Psi}_o - \sigma_e \tilde{\Psi}_e) + (\tilde{\Psi}_o - \tilde{\Psi}_e) \right]$$ (17)

The obtained expressions describe oblique propagation of a fundamental vortex-beam in a uniaxial crystal.

## III. The dislocation reactions in the field components

### III.1 The energy flux

The equations (16) and (17) describe the intrinsic features of the phase and polarization singularities in the oblique beams in a crystal. Each component of the field comprises a superposition of two individual beams (the ordinary and extraordinary ones) transmitted along different directions defined by the angles $\alpha_o$ and $\alpha_e$ (see Fig.1).

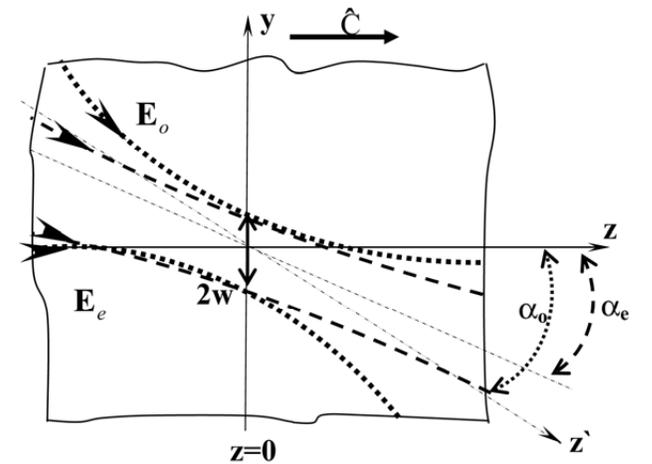

Fig.1 The sketched representation of the off-axis individual paraxial beams in the uniaxial anisotropic medium

Besides, these individual beams have different structure of the wavefront defined by the wavenumbers $k_o$ and $k_e$. The immediate corollary of such field behavior is that the energy flux in each field component is quickly oscillated at any crystal section via variations of the beam parameters. Indeed, z-component $P_z$ of the energy flux of the beam field in the paraxial case can be calculated as:

$$P_z \propto \int\int_{-\infty}^{\infty}\left(|E_+|^2+|E_-|^2\right)dx\,dy = I_+ + I_-. \quad (18)$$

By using eqs.(16) and (17) we find the dependency of the energy flux $P_z$ in each field component on the angle $\alpha_o$ shown in Fig.2. When the distance between two maxima in the interferential pattern is larger than the waist radius of the beam and the dark interferential fringe is positioned at the center of the beam cross-section we observe the intensity minimum. The lesser is the waist radius $w_o$ the larger number of interferential fringes are positioned on the beam cross-section. The intensity oscillations are smoothed (compare Fig.2, a and b). While increasing the angle $\alpha_o$ the oscillations are smoothed too.

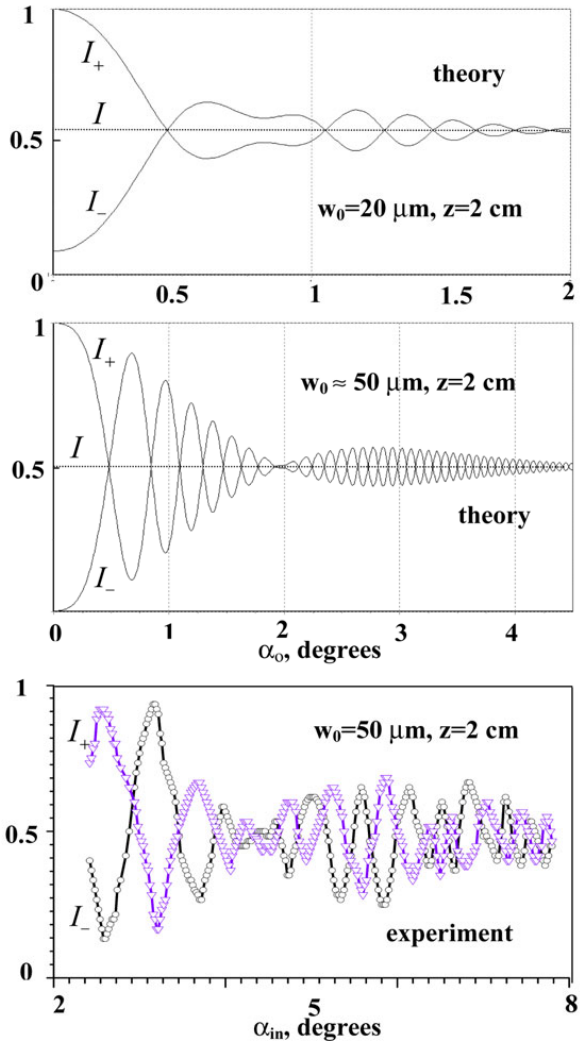

Fig.2 Variations of the total beam intensities $I_+$ and $I_-$ for $E_+$ and $E_-$ components of singular beam in LiNbO$_3$ crystal with $n_o = \sqrt{\varepsilon} = 2$, $n_o = \sqrt{\varepsilon_3} = 2.1$ for different beam waists $w_0$.

We investigated the oscillations of the total intensity with the help of the experimental set-up whose sketch is shown in Fig.3. The Gaussian beam emitted from He-Ne laser $(\lambda = 0.6328\,mcm)$ passing through the polarization filter (the polarizer $P_1$ and λ/4 plate) is transformed into a right-handed circularly polarized one. The optical wedge $W$ turns the Gaussian beam into a singular one with a centered vortex having a topological charge $l = -1$. The diaphragm $D_1$ permits to regulate the radius of the beam waist $w_0$ at the crystal input. The lens $L_2$ with $f=3$ cm focuses the beam onto the input face of the LiNbO$_3$ crystal with the optical axis directed along the z-axis (see Fig.3a). The crystal length is $z \approx 2$ cm, the refractive indices for the major crystal directions are about $n_1 = n_o = \sqrt{\varepsilon} \approx 2.2$, $n_3 = \sqrt{\varepsilon_3} \approx 2.3$. The crystal is mounted on the special rotary table. The angle accuracy of the table rotation is about 0.03°. The beam behind the crystal is collimated by the lens $L_3$. The quarter-wave plate λ/4 and the polarizer $P_2$ can filter off either the left-hand $E_-$ or right-hand circularly polarized $E_+$ components so that we can observe the intensity distribution of the $I_+$ and $I_-$ components on the computer monitor. We can also observe interference patterns of the field at the crystal output. This is reached up by means of the Mach-Zender interferometer. We change the incident angle $\alpha_{in}$ of the beam on the crystal input and observe the intensity distribution on the monitor. The set-up permits to measure also the total intensity of the field in relative units.

Typical intensity oscillations as a function of the incident angle $\alpha_{in}$ of the beam at the crystal input face $z=0$ illustrate Fig.2. The incident angle $\alpha_{in}$ and the inclination angle $\alpha_o$ of the ordinary beam in the crystal are connected with each other by a simple relation $n_0 \alpha_{in} \approx n_o \alpha_o$ (where $n_0 = 1$). While changing the radius $w_d$ we can smoothly change the radius of the beam at the crystal input. The optimal results in our experiment were reached for the beam radius $w_o \approx 50 \mu m$. It is such a value of the radius $w_o$ and the utmost accuracy of the angle variation $\Delta\alpha_{in} \approx 0.03^o$ that enables us to plot the relatively smooth experimental curves $I_{\pm}(\alpha_{in})$ presented in Fig.2. Notice that the total intensity $I = I_+ + I_-$ experiences no variations via the angle $\alpha_{in}$ but its value depends on the beam waist $w_o$.

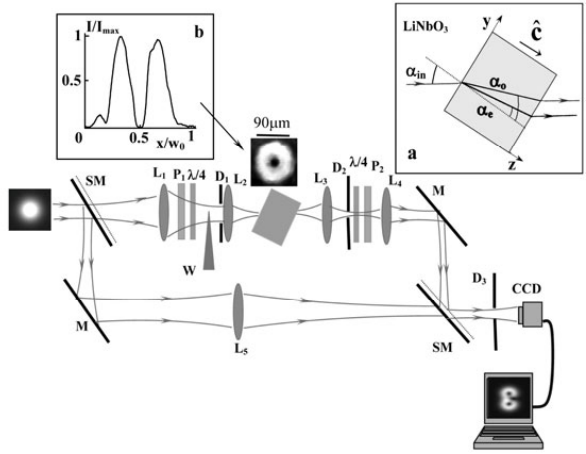

Fig.3 Sketch of the experimental set-up: $P_1, P_2$-polarizers, λ/4 - quarter wave plate, $L_1$-$L_5$-lenses, W-optical wedge, $D_1$-$D_3$ - diaphragms, SM - semi-transparent mirrors, M – mirrors, CCD - CCD camera; (a) Beam axis trajectories in LiNbO$_3$ crystal $\hat{\mathbf{c}}$ - a unit vector of the crystal optical axis;

Our preliminary measurements showed also that 1) initial optical vortex splits in the crystal into four centered optical vortices imprinted in four individual beams; 2) a right-handed $E_+$ and left-handed $E_-$ components of the electric vector carry over opposite-charged optical vortices on the axis at very small angles $\alpha_{in} \approx 0$; 3) after splitting the individual beams at the angles $\alpha_{in} > 10^o$ all four centered vortices have the same topological charges. Thus, hereafter we will focus our attention on bringing to light a physical mechanism responsible for such a conversion process.

**III.2 Dislocation reactions**

While a superposition of two oblique non-singular beams results in the interferential pattern in the form of ordinary fringes or rings, the interference of two oblique singular beams entails the dislocation reactions: chains of birth and annihilation events of optical vortices. In general case, any variations of the angle $\alpha_o$ cause deformations of the beam structure consisting of polarization singularities: C-lines and L-surfaces [18]. L-surface represents a locus where the electric field is linearly polarized. C-line forms a locus where one of the circularly polarized field components vanishes. Its amplitude is zero while the phase is uncertain. L-surfaces encircle C-lines. In fact, C-line represents a space trajectory of the optical vortex imprinted in the circularly polarized field component. In this Section we will bring to light intrinsic features of the dislocation reactions accompanying the deformation of the beam structure in one of the circularly polarized components.

At first, let us consider transformations of phase singularities in the $E_-$ component of the oblique beam (eq. (17)) provided that only the $E_+$ component is at the plane $z=0$:. We can imagine that the input face of the crystal coincides with the plane $z=0$ and consider the demand: $E_-(z=0) = 0$ while the reflected beam is neglected as the boundary condition.

Let the right-hand circularly polarized beam bearing the vortex with a negative topological charge ($l = -1$) falls on the crystal input perpendicular to its verge ($\alpha_{in} = \alpha_o = 0$). Equation (17) shows the left-hand circularly polarized component $E_-$ in the crystal to carry over the on-axis optical vortex with a positive topological charge $l = +1$. Fig.4 illustrates the theoretically predicted and experimentally measured intensity distributions and interferential patterns of the $E_-$ component. The one-branched spiral with a left-handed rotation for $\alpha_{in} = 0$ is evidence of the centered optical vortices with a positive topological charge (the spiral in the $E_+$ component is twisted in the opposite direction relative to that in the $E_-$ component). When growing the angle up to $\alpha_{in} = 0.5^o$ a double-charged vortex is torn off the beam following the crystal optical axis while a singly-charged vortex with $l=-1$ follow the beam direction. At the experiment we observe the interference pattern in the form of a two-branched spiral. As the angle $\alpha_{in}$ increases ($\alpha_{in} > 2^o$) the portion of the light intensity falling on this vortex essentially decreases so that both the $E_+$ and $E_-$ components of the oblique beam carry over the vortices near their axes *with the same signs of the*

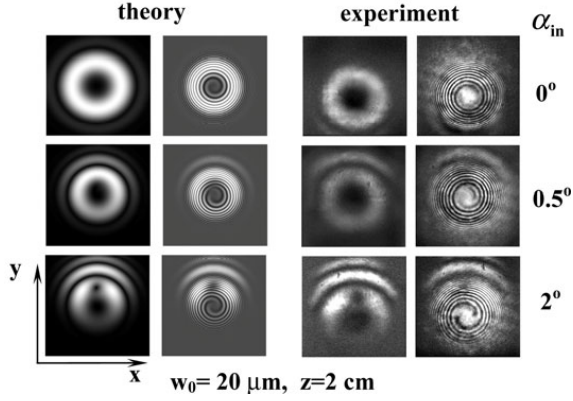

Fig 4 Theoretically predicted and experimentally measured intensity distributions and interferential patterns of $E_-$ component of the off-axis beam via different incident angle $\alpha_{in}$

*topological charge* ($l = -1$) *equal to that in the initial one at the crystal input*. In order to understand a fine structure of the beam we appealed to vortex trajectories, written as

$$\mathrm{Re}\left[E_\pm(x,y,z,\alpha_o)\right]=0, \ \mathrm{Im}\left[E_\pm(x,y,z,\alpha_o)\right]=0. \quad (19)$$

The structure of the projections of these trajectories onto the plane $x0y$ at the starting range: $0 \leq \alpha_o \leq 1^o$ is shown in Fig.5a. The red and black thick points in the figure refer to the birth and annihilation events, respectively. The white ringlet indicates the initial vortex. The sequence of the ciphers characterizes the succession of the birth and annihilation events.

Let us consider dislocation reactions in the $E_-$ component. In order to observe some vortex processes with better resolution than that in Fig.4 we choose an initial singular beam with a relatively large waist radius. Fig.5 shows that no sooner the angle $\alpha_o$ has started to change (event 1) than the positively charged optical vortex ($l = +1$) slid off the beam axis. Then the topological dipole is born near the beam axis $x=0$, $y=-\alpha_o z$ (event 2, red point). The positively charged vortex of the dipole pair aspires to the initial vortex far from the $z$-axis (Fig.5a) whereas the negatively charged one follows the beam forming the general branch of the trajectory. Two positively charged vortices draw together at some distance from z-axis. Their structure looks like an entire double-charged vortex in the interferential experiment with a relatively small waist of the initial beam (see Fig.4). However, the vortices don't flow together drifting very slowly along $x$-direction (Fig.5a). Two dipole pairs are born again at the points 3' and 3'' almost simultaneously. Two opposite-charged vortices of these pairs annihilate at the point 4 (black point) while the rest two identically charged vortices scatter very quickly in opposite directions forming one of the transverse branches of the complex trajectory. On the general branch at the point 5, a new vortex pair is born. Two opposite-charged vortices annihilate at the point 6 while the residuary negatively charged vortex continues to build the general trajectory branch. Later on this succession of the events is reproduced right up to the critical angle $\alpha_{cr}$ where the general branch is bisected (see the next Section). The dislocation reactions in the $E_+$ component occur just as in the $E_-$ component if the very starting range of the trajectory is not taking into account. Variations of the beam waist $w_0$ at the crystal input entail the deformation of the trajectory but its intrinsic features are preserved. Notice that some basic properties of this complex vector processes were perceived on the example of a simple scalar model of composite vortices [19] comprising two decentered singly charged singular beams whose axes are shifted parallel to each other and the beam fields have some phase difference.

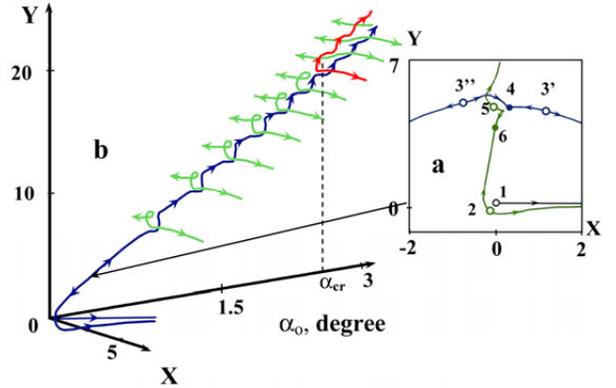

Fig.5 Trajectory of optical vortices imprinted in the $E_-$ component,: a) the region of the vortex trajectory in X,Y coordinates for the LiNbO3 crystal inside the range of angles $\Delta\alpha_o$ $(0,1^o)$ b) the vortex trajectory in XY$\alpha$ cordinates, the crystal length z=2 cm

## IV. The indistinguishability limit

As we have seen above any variations of beam parameters entail birth and annihilation vortices. We observe a complex vortex mixture. However, the experiment showed that starting with some critical angle $\alpha_{cr}$ (or some crystal length $d_{cr}$) there are two stable vortices in each circularly polarized component taking no part in dislocation reactions.

Each of these vortices is a special mark of an individual beam in the structure of the combine beam that follows the individual beam while changing parameters of the initial beam. The event when four stable vortices appear in the beam component is a distinguishability limit for the beam parameters starting with which the individual beams can be observed separately from the other one [20]. As a whole this limit characterizes quadrefringence of the initial vortex in a uniaxial crystal: splitting the initial vortex into four stable, centered ones.

However, *in order to distinguish two beams in each circularly polarized component it is necessary, at first, to define which singular beam the individual vortex belong to*. The important role in this process play the portion of the vortex trajectory in vicinity of the critical angle $\alpha_{cr}$ in Fig.5b. The major branch of the vortex trajectory are divided here and do not interlace any more. Notice that the total number of vortices to be observed simultaneously on all trajectory branches for $\alpha < \alpha_{cr}$ can be varied from one to four. While transiting the border $\alpha = \alpha_{cr}$ one optical vortex disappears. It is this critical angle that characterizes the beam splitting. The birth and annihilation events emerge at the sidelobes of the beams forming the transversal trajectories. The individual beams can be distinguished. A total number of vortices observed simultaneously beyond the angle $\alpha > \alpha_{cr}$ is varied from two to four. The critical angle $\overline{\alpha} = \overline{\alpha}_{cr}$ can be regarded as the indistinguishability limit for two refracted beams with the same polarizations.

We can estimate mathematically the value of the critical angle $\alpha_{cr}$ based on eq. (16) for $E_+$ component. Indeed, the optical vortices in two individual beams (the first and the second terms in eq. (16)) take part in no dislocation reactions when the front intensity maximum of the first individual beam coincides with the back intensity maximum of the second individual one. By using eq. (16) we come to the expressions for the coordinates of the intensity maximum $y_{front}^{(o)} = \alpha_o z + w_o / \sqrt{2}$, $y_{back}^{(e)} = \alpha_e z - w_e / \sqrt{2}$, where $w_o = w_0 \sqrt{1 + z^2 / z_o^2}$, $w_e = w_0 \sqrt{1 + z^2 / z_e^2}$. The maxima matching obey the condition $\Delta y = y_{front}^{(o)} - y_{back}^{(e)} = 0$ from whence we find for $z_o / z \ll 1$:

$$z^2 \left( \frac{\alpha_o^2}{w_0^2} \frac{\Delta\varepsilon^2}{2\varepsilon_3^2} - \frac{1}{z_o^2} \right) \approx 1 \qquad (20)$$

The last equation shows that we cannot distinguish two singular beams at any crystal length $z$ if the angle of the ordinary beam is less than $\alpha_o = \alpha_{cr} = \varepsilon_3 \sqrt{2\alpha_{diff}} / |\Delta\varepsilon|$ where $\alpha_{diff} = 2 / k_o w_0$.

The curve presented in Fig.6a characterizes the indistinguishability limit for two singular beams. The magnitudes $(\alpha_o, w_0)$ located on the right of the curve are associated with two beams that can be distinguished at the experiment. The non-distinguishable fields are perceived as one composite beam. The variation of the angle $\alpha_o$ is accompanied by the appearance of the phase singularities: from one to four vortices at the beam cross-section in the indistinguishability area and from two to four vortices in the distinguishability area. As we have said above (see Sec.III), the total intensity of the field component $E_+(\alpha_o)$ oscillates sharply within the range $(0, \alpha_{cr})$. The oscillation is smoothed, i.e, the amplitude of the oscillations decreases and the period increases beyond the critical angle $\alpha_{cr}$. The intensity maxima correspond to the minimum number of vortices (one or two) in the beam. The vortices on the transversal trajectories leave very quickly the area of the steady observation (at the experiment). The angle interval between two maxima is approximately equal to $\Delta\alpha \approx \alpha_o |1 - \varepsilon_o / \varepsilon_3|$. For a typical case, $n_o = 2.3$, $n_3 = 2.208$ $\alpha_o \approx 1.8^o$ we have $\Delta\alpha \approx 0.15^o$. Thus, within the region $\Delta\alpha$ we can register one or two vortices in the indistinguishability range and two vortices in the distinguishability range. The rest vortices leave the observation area very quickly via slight variations of the angle $\alpha_o$.

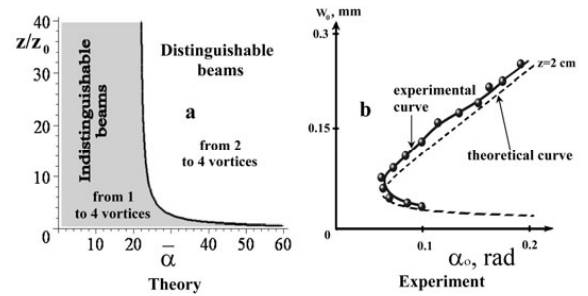

Fig 6 The curves $\alpha_o = f(\rho, z)$ outlining the indistinguishability range for two singular beams

These reasons formed the basis for our experimental measurements of the indistinguishability limit. Besides, it was experimentally insufficient to use only one lens presented in Fig.3 with the iris diaphragm for forming relatively large waist radii of the beams at the crystal input. We employed three lenses with the focus distances $f = 5 cm$ $f = 12 cm$ and $f = 20\ cm$. Presence of the vortices in the beam and the signs of their topological charges were measured by means of the interferometric methods. We measured the indistinguishability limit $\alpha_{cr}$ with accuracy about $\delta\alpha \approx 0.1^o$. It makes up $\alpha_{cr} \approx 2.7^o$ inside the crystal for the beam waist at the crystal input $w_0 \approx 50 mcm$ and the crystal length $z = 2 cm$. Notice that the diaphragm $D_1$ in Fig. 3 distorted to some extent a shape of the initial Gaussian envelope at the crystal input in the form of Airy rings. Such distortion increases when decreasing the diaphragm pupil and the beam waist. The experimental error increases. Nevertheless, the results shown in Fig.6b illustrate a good agreement with theoretical predictions.

## V. The vortex quadrefringence

Let us focus our attention on splitting of a centered vortex at the plane z=0 into four centered vortices bearing four individual beams in terms of polarization singularities [18]. The vector singularities of off-axis beams in the uniaxial crystal were considered recently [13, 14] within a small range of the angles for the KDP crystal and a very large input beam waist ($w_0 \approx 0.7\ mm$) on the base of a reductive model of the beam propagation in the birefringent media when the initial beam is a linearly polarized. This regime corresponds to the indistinguishability range (the refracted beams in each polarized component cannot be distinguished). They revealed very complex structure of C-lines resembling the braided vortex trajectories besides, different C-line branches can be reconnected via variations of the angle $\alpha_{in}$ (or the phase difference between the beams with orthogonal linear polarizations). In the given Section we will consider pattern as a whole for a broad range of the angles $\alpha_o$.

As well-known [18], vortex trajectories in each circular polarized component and C-lines in a nonuniformly polarized paraxial beam are tightly connected with each other. We made use of such conformity and plotted C-lines without calculation of Stokes parameters. The computer simulation of the evolution of C-lines via variations of the angle $\alpha_o$ is depicted in Fig.7. The curves in the figure can be divided into two groups (green and blue) associated with the vortex trajectories in the $E_+$ and $E_-$ field components. Besides, inside the group we can speak about C-lines of the general and transversal branches. The lines of all branches are neither intersected nor reconnected although reconnections emerge along each branch within the indistinguishability range. The exceptions are only for the starting sections of the new-born branches of C-lines in the splintered beams near the angles $\alpha_{cr}^{(+)}$ and $\alpha_{cr}^{(-)}$ when the transversal and general branches are united. Each two lines of two different groups (red and blue) behind the critical angles $\alpha_{cr}^{(+)}$ and $\alpha_{cr}^{(-)}$ form two distinct braidlets stretched along the general direction. Notice that the shape of C-lines in our case has much to do with that presented in Ref. 13 and 14 for the angle range $\alpha < \alpha_{cr}$.

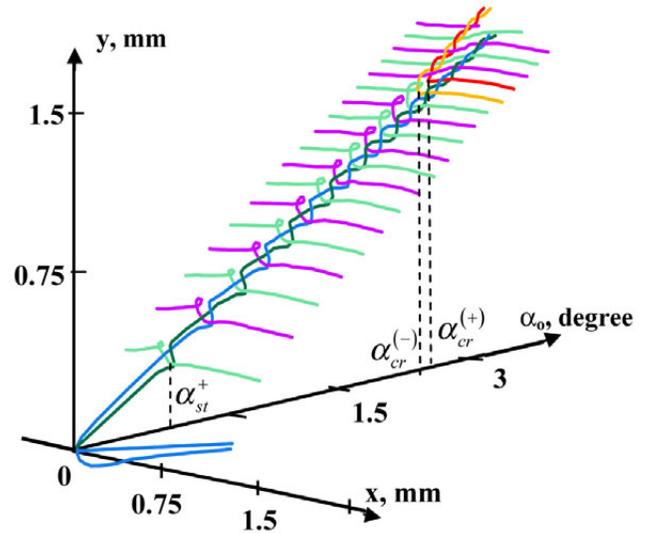

Fig. 7  C-lines for the singular beam with $w_0 = 50\ \mu m$, z=2 cm

The important point in our consideration is that the branches in each braidlet do not interflow after splitting the lines at the critical angles for any crystal lengths. Indeed, let us consider in eqs (16, 17) that the magnitude $\overline{\alpha}\ Z \gg 1$, where $Z_o = z/z_o$, $\overline{\alpha} = \alpha_o / \alpha_{diff}$. A simple algebra shows that in eq. (17): $[(x + i\ y_o)/r_o]^2 \approx -(1 - 2i\ x/\overline{\alpha}Z)$, where we restrict ourselves to the first order of smallness for the value $1/\overline{\alpha}Z$.

Finally, we come to the expressions:
$$\widetilde{E}_+ \propto \frac{X - i Y'_o}{Z_o} \widetilde{\Psi}_o + \frac{X - i Y'_e}{Z_e} \widetilde{\Psi}_e,$$

$$\tilde{E}_- \propto \frac{(X+1/\overline{\alpha})-iY_o}{Z_o}\tilde{\Psi}_o - \frac{(X+1/\overline{\alpha})-iY_e}{Z_e}\tilde{\Psi}_e, \quad(21)$$

where $Z_e = z/z_e$, $X = x/w_0$, $Y_o' = (y-\alpha_o z)/w_0$, $Y_e' = (y-\alpha_e z)/w_0$. First of all, we see that vortices have the same signs of the topological charges in both circularly polarized components. Besides, we find from eqs (21) that four nodal lines for $E_+$ and $E_-$ components lie on the plane $Y_o = 0$ and $Y_e = 0$ planes for $\overline{\alpha}Z \gg 1$. Coordinates of the vortices on the neighbouring trajectories differ in the value:

$$\Delta x = w_0/\overline{\alpha} = \frac{\lambda}{\pi n_o \alpha_o} = \frac{\lambda}{\pi n_e \alpha_e}.$$

where $\lambda$ stands for a wavelength in free space. For example, for the initial beam with $w_0 = 50\,\mu m$, $\alpha_o = 5^o$, $n_o = 2.3$ the value of the trajectory splitting is $\Delta x \approx 1\,\mu m$. It means that in the frames of our approximation, nearly centered vortices bearing four individual beams do not flow together after splitting for any crystal lengths.

Using the method presented in Ref. 10 we plotted the maps of polarization states for a sufficiently large angle $\alpha_o = 5.3^o > \alpha_{cr}$ shown in Fig.8a,b. The individual beams are practically separated. The uniform field of the linear polarization covers over the most part of the beam cross-sections similar to that presented in the model in Ref. 13 and 14. Only a thin stria of the nonuniform polarization states parts the individual beams. However, the pairs of singular points in the form of the lemon and the star are positioned inside a dark area of each beam perturbing a usual uniform picture of the birefringent process (see Fig.8b). We examined this singular structure with the help of a computer simulation for a broad range of the crystal lengths (right up to z=20 cm). We revealed that there are two C-points encircled by the pattern in the form of the star and the lemon in vicinity of minimum of each individual beam. They approach to each other at the distance about $1\,\mu m$ without a polarization unfolding at the crystal length z=20 cm that is a good agreement with our estimation presented above. The map shown in Fig.8c,d pictures the polarization states and the characteristic integral lines obtained at the experiment. We employ the computer-processing technique for plotting the map described in Ref.10 and 21. This technique enabled us to measure all Stokes parameters using only four measurements instead of six ones. It reduced sufficiently an exponential error and increased accuracy of the experimental plottings. Besides, our CCD camera has 160 pixel per mm that permits to resolve the polarization states in the map with accuracy not worse than 1μm.

Fig.8c,d describes the composite beam for the angle $\alpha_{in} = 12^o$ (or $\alpha_o \approx 5.2^o$) at the crystal input. The nonuniform distribution of polarization states at the area of the dark spots in the Fig.8d contrasts sharply with a nearly uniform distribution of the linear polarization state stretched over the rest potion of the beam cross-section in Fig.8c. Two singular points: the lemon and the star (corresponding to two optical vortices in the field components) are positioned at the distance about 20 μm. This value is larger at one order of magnitude

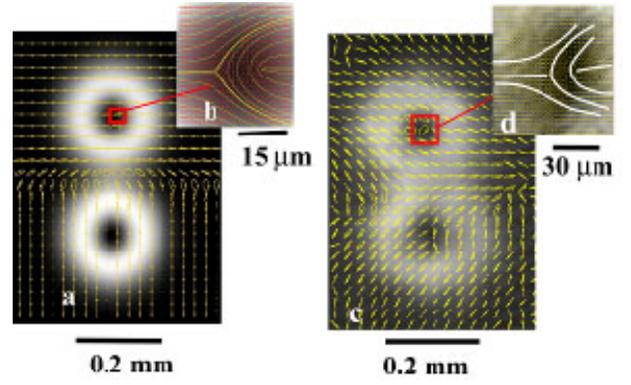

Fig.8 The maps of polarization states and integral curves for directions of the azimuthal angle of the polarization ellipses for the angles $\alpha_o$=5.3 ° ($w_0$=50 μm, z=2 cm) positioned against a background of the intensity distributions: (a,b) –theory, (c,d) - experiment

than our theoretical estimation $\Delta x \approx 1\,\mu m$. Such disagreement is caused by the fact that the value $\overline{\alpha}Z_o \approx 0.36$ in our experiment does not corresponds to the condition $\overline{\alpha}Z_o \gg 1$. At the same time, this experimental result is in a good agreement with a computer simulation in Fig.8b.

As a whole, the initial vortex at the plane z=0 splits at first into two nearly centered vortices in the orthogonal field components (two C-lines in Fig.7 for very small angle $\alpha_o$). Then after a critical angles $\alpha_{cr}^{(+)}$ and $\alpha_{cr}^{(-)}$ there appear four nearly centered C-lines associated with four optical vortices. *It looks like as if the initial vortex splits into four ones, i.e. it experiences a quadrefringence.*

### VI. Conclusions

We have found the solutions to the paraxial wave equation in a uniaxial anisotropic medium in the form of oblique fundamental vortex-beams. We have theoretically and experimentally analyzed the singular structure of the beams plotting the vortex trajectories and mapping polarization states for different propagation directions of the initial beam. We revealed that centered initial optical vortex at the crystal input splits into two ones in each circularly polarized component. These vortices take part in no dislocation reactions and their trajectories do not coincide in space for any crystal lengths. We called this process the vortex quadrefringence.

## ACKNOWLEDGEMENT

Authors thank A. P. Kiseleev for useful discussion and K. Kotlyarov for the help in the experiment.